\documentclass{ws-ijmpa}

\begin{document}

\markboth{R. Foot}{Experimental Implications of mirror matter-type dark
matter}

\catchline{}{}{}{}{}

\title{Experimental implications of mirror matter-type dark matter}

\author{R. Foot}
\address{
rfoot@unimelb.edu.au \\
School of Physics, University of Melbourne, 3010 Australia. }


\maketitle


\begin{abstract}
Mirror matter-type dark matter is one dark matter candidate
which is particularly well motivated from high energy
physics. The theoretical motivation and experimental evidence are 
pedagogically reviewed, with emphasis on the implications of recent 
orthopositronium
experiments, the DAMA/NaI dark matter search, anomalous 
meteorite events etc.

\end{abstract}



\vskip 0.6cm

There is very strong evidence that the Universe has 
a large non-baryonic dark matter component. On the other hand,
the standard
model of particle physics does not contain any 
heavy, stable non-baryonic particles. Clearly,
this motivates new particle physics beyond the standard
model.

It seems to me that the most interesting candidate for
this new physics is mirror symmetry.
It is the most interesting candidate because it involves
only a single well motivated hypothesis. 
Parity and time reversal symmetries stand out as the 
only obvious symmetries which are not respected by
the interactions of the known elementary particles. It is an interesting
and non-trivial fact that these symmetries can
be exact, unbroken symmetries of nature if a 
set of mirror particles exist. 
Even more interesting is that the mirror particles
have the right broad properties to be identified with the non-baryonic
dark matter in the Universe.
But we are running ahead of ourselves. Let us start
at the beginning...

In 1956 Lee and Yang\cite{ly6} proposed that the interactions
of the fundamental particles were not mirror reflection
invariant. They suggested that this could explain
some known puzzles and proposed some new experiments 
to directly test the idea. Subsequently Madam C.S.Wu and
collaborators dramatically confirmed that the interactions
of the known particles were not mirror symmetric, just as
Lee and Yang had suspected.

Today, it is widely believed that mirror symmetry is in fact
violated in nature. God -- it is believed -- is left-handed.
Actually, though, things are not so clear.
What the experiments in 1957 and subsequent
experiments have conclusively demonstrated is that
the {\it known} elementary particles behave in a way which
is not mirror symmetric. 
The weak nuclear interaction is the culprit, with the
asymmetry being particularly striking for the weakly
interacting neutrinos.
For example, today we
know that neutrinos only spin with one orientation. If one
was coming towards you it would be spinning like a left-handed
corkscrew. Nobody has ever seen a right-handed neutrino.

The basic geometric point is illustrated in the following diagram:
\vskip 0.4cm
\centerline{\epsfig{file=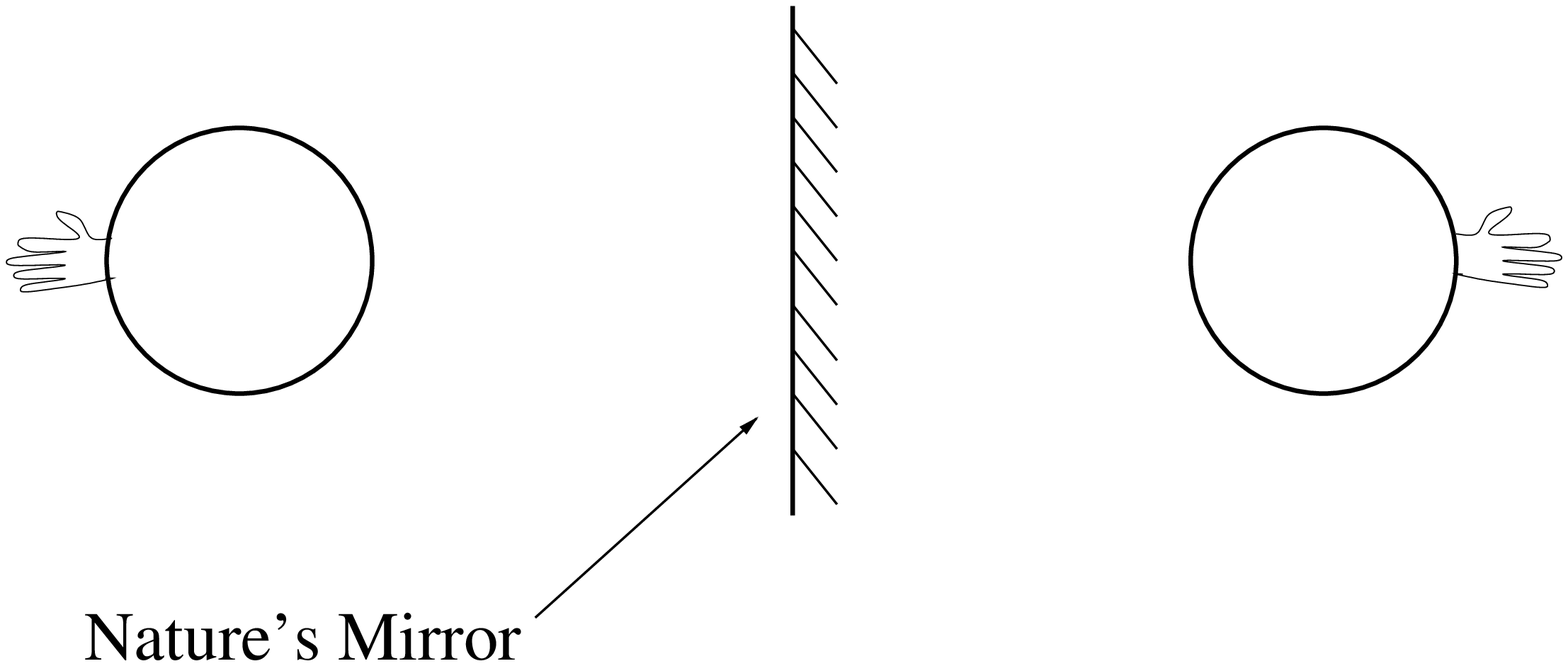,width=7.1cm}}
\vskip 0.4cm
\noindent 
The left-hand side of this figure represents the interactions
of the known elementary particles. The forces are
mirror symmetric like a perfect sphere, except for
the weak interaction, which is represented as
a left hand. Also shown is nature's mirror - the vertical
line down the middle. Clearly, the reflection is not the same as
the original, signifying the fact that the 
interactions of the {\it known} particles are not mirror symmetric.
If there were a right hand as well
as a left hand then mirror symmetry would be
unbroken.
\vskip 0.4cm
\centerline{\epsfig{file=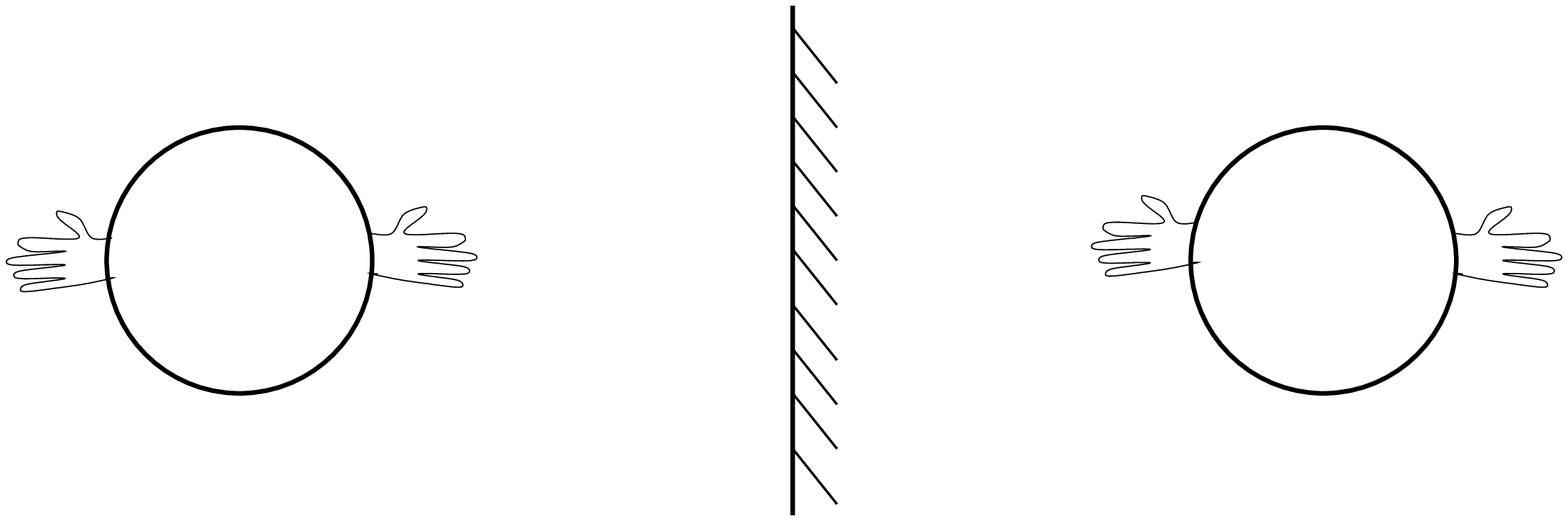,width=7.0cm}}
\vskip 0.4cm
\noindent
However, this doesn't correspond to nature since no right-handed
weak interactions are seen in experiments (this is precisely what the 
experiments in 1957 and subsequently have proven).

There are two remaining
possibilities: We can either chop the hand off -- but this is too
violent and is therefore not shown. It corresponds to having no weak 
interactions at all, again in disagreement with observations.
This last possibility is the most subtle and consists
of adding an entire new figure with the
hand on the other side. Everything is doubled even the symmetric part,
which is clearly mirror symmetric as indicated in
the following diagram:
\vskip 0.4cm
\centerline{\epsfig{file=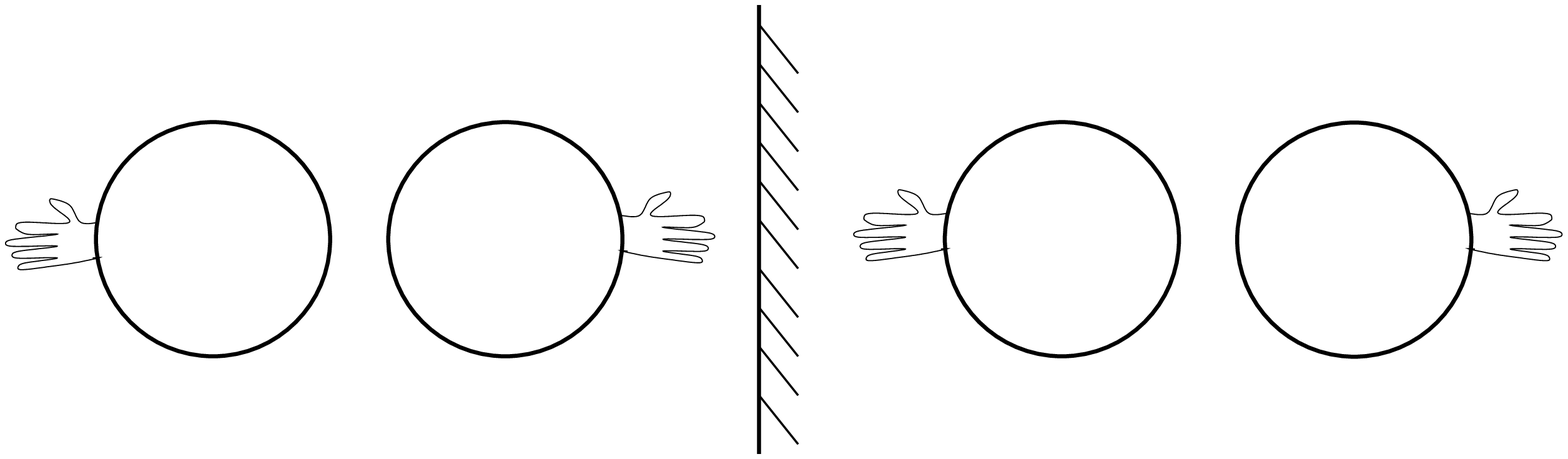,width=8.2cm}}
\vskip 0.4cm
\noindent 
What this figure corresponds to is a complete doubling 
of the number of particles. For each type of particle, such
as electron, proton and photon, there is a mirror twin. 
Where the ordinary particles favor the left hand, the
mirror particles favor the right hand. If such particles
exist in nature, then mirror symmetry would be exactly conserved
(we denote the mirror particles with a prime).
\vskip 0.2cm
\centerline{\epsfig{file=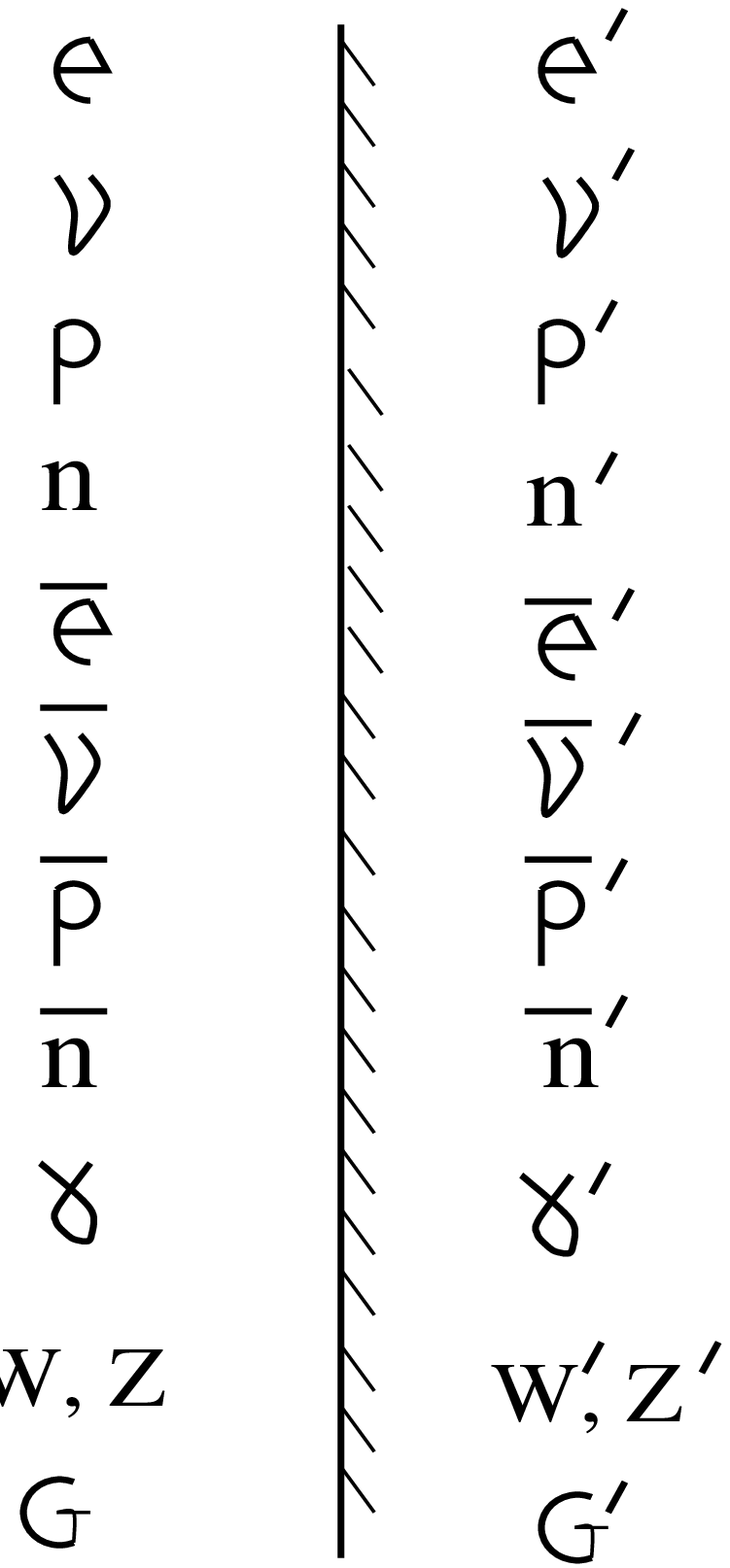,height=5.2cm,width=2.6cm}}
\vskip 0.2cm
\noindent
As will be discussed, the mirror particles can
exist without violating any known experiment. 
Thus, the correct statement is that the experiments in
1957 and subsequently have only shown that the interactions
of the {\it known} particles are not mirror symmetric, they
have not demonstrated that mirror symmetry is broken in nature.

The ordinary and mirror particles form parallel sectors each
with gauge symmetry $G$ (where $G=G_{SM} \equiv SU(3)_c \otimes SU(2)_L 
\otimes U(1)_Y$
in the simplest case) so that the full gauge group is $G \otimes G$.
Mathematically, mirror symmetry has the form:\cite{flv}
\begin{eqnarray}
& x \to -x, \ t \to t, \nonumber \\
& W^{\mu} \leftrightarrow W'_{\mu}, \ B^{\mu} \leftrightarrow B'_{\mu},
\ G^{\mu} \leftrightarrow G'_{\mu} \nonumber \\
& \ell_{iL} \leftrightarrow \gamma_0 \ell'_{iR}, \
e_{iR} \leftrightarrow \gamma_0 e'_{iL}, \
q_{iL} \leftrightarrow \gamma_0 q'_{iR}, \
u_{iR} \leftrightarrow \gamma_0 u'_{iL}, \
d_{iR} \leftrightarrow \gamma_0 d'_{iL}, 
\end{eqnarray}
where $G^{\mu}, W^{\mu}, B^{\mu}$ are the standard
$G_{SM} \equiv SU(3)_c \otimes SU(2)_L \otimes U(1)_Y$ gauge particles,
$\ell_{iL}, e_{iR}, q_{iL}, u_{iR}, d_{iR}$ are the
standard leptons and quarks ($i=1,2,3$ is the generation index)
and the primes denote the mirror particles. There is also a 
standard Higgs doublet $\phi$ with a mirror Higgs doublet partner,
$\phi'$, and it can be shown that $\langle \phi \rangle =
\langle \phi' \rangle$ for a large range of parameters of the 
Higgs potential\cite{flv}. 
This means that the mirror symmetry is {\it not} spontaneously
broken by the vacuum, so that it is an exact, unbroken
symmetry of the theory.
Interestingly, despite doubling the
number of particle types the number of free parameters have
not (yet!) been increased: mirror symmetry implies that the masses and 
couplings of the 
particles in the mirror sector are exactly the same as the 
corresponding ones in the ordinary sector.

Ordinary and mirror particles couple with each other via gravity
and possibly by new interactions connecting ordinary and mirror
particles together. 
Constraints from gauge invariance, mirror symmetry and
renormalizability, suggest only two
types of new interactions\cite{flv}:\footnote{
Allowing the ordinary and mirror sectors to interact with each
other leads to 
two new free parameters ($\lambda', \epsilon$).
However, compared to
other ideas beyond the standard model, many of which have
literally hundreds of new parameters, mirror symmetry {\it is}
a fairly minimal extension of the standard model.
Also note, if the neutrinos have mass, mass mixing between ordinary and
mirror neutrinos is also possible\cite{flv2,f94} and might be
implicated by the observed atmospheric, solar and 
LSND neutrino anomalies. 
However, the experimental situation is still not clear\cite{foot}.
}
a) Higgs-mirror Higgs quartic coupling
(${\cal L} = \lambda' \phi'^{\dagger}\phi' \phi^{\dagger} \phi$),
and b) via photon-mirror photon kinetic mixing:
\footnote{
Technically, the photon-mirror photon kinetic mixing arises from
kinetic mixing of $U(1)_Y, U(1)'_Y$ gauge 
bosons, since only for abelian $U(1)$ symmetry is the mixing
term, $FF'$, gauge invariant.
Therefore there is both $\gamma-\gamma'$ and  $Z-Z'$
kinetic mixing. [However, experiments are much more sensitive to
$\gamma - \gamma'$ kinetic mixing which is why we focus
attention on it].
In the case of theories without $U(1)$ gauge symmetries, such
as GUTs, the $\gamma-\gamma'$ mixing can arise radiatively\cite{hol}.
Interestingly, there is a class of models where $\epsilon$
vanishes at one and two loop level\cite{cf}, and therefore
naturally of the order of $\epsilon \sim 10^{-8}$.} 
\begin{eqnarray}
{\cal L}_{int} = {\epsilon \over 2}F^{\mu \nu}F_{\mu \nu}' .
\label{km}
\end{eqnarray}
where $F^{\mu \nu}$ ($F'_{\mu \nu}$)
is the field strength tensor for electromagnetism (mirror
electromagnetism).
The effect of the Higgs-mirror Higgs quartic coupling is to
modify the properties of the standard Higgs boson\cite{flv,flv2,sasha}. This
interaction will be tested if/when scalar particles are
discovered. The effect of photon-mirror photon kinetic mixing is to
cause mirror charged particles
to couple to ordinary photons with effective electric
charge $\epsilon e$\cite{flv,hol,s}.

This leads to a number of very interesting effects. 
In the laboratory, mirror particles can potentially be
produced from interactions of ordinary particles:
$e^+ e^- \rightarrow e'^+ e'^-$. 
The Feynman diagram is given in the following figure:
\vskip 0.3cm
\centerline{\epsfig{file=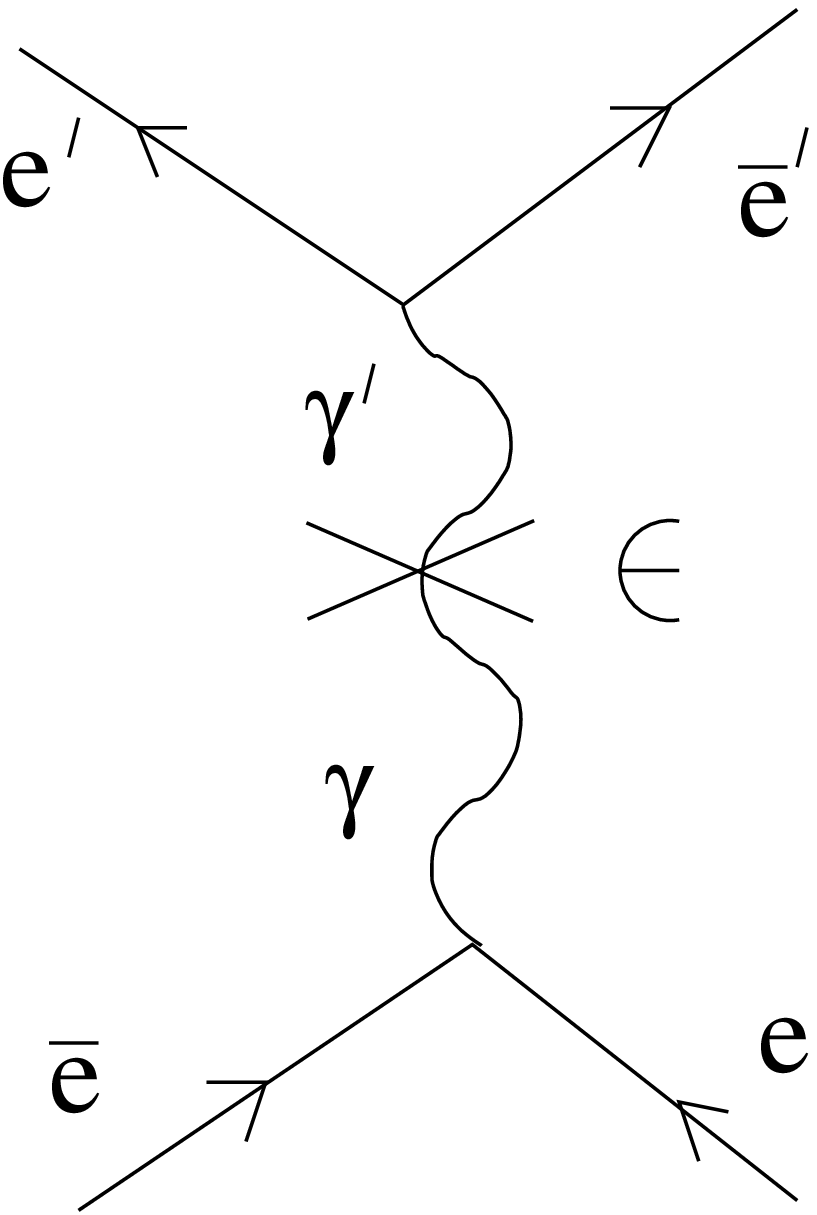,height=4.7cm, width=3.1cm}}
\vskip 0.50cm
\noindent
The best laboratory limits
for the production of such light stable ``minicharged'' particles
comes from the SLAC beam dump experiment\cite{slac},
$|\epsilon | \stackrel{<}{\sim} 10^{-4}$.
However, this is not the most sensitive laboratory
test. A more sensitive laboratory test for mirror
matter comes from the orthopositronium
system\cite{gl}. The interaction of $e^+ e^-$ with $e'^+ e'^-$
leads to a small mass term mixing orthopositronium with
mirror orthopositronium. 
The Feynman diagram is given in the following figure.
\vskip 0.4cm
\centerline{\epsfig{file=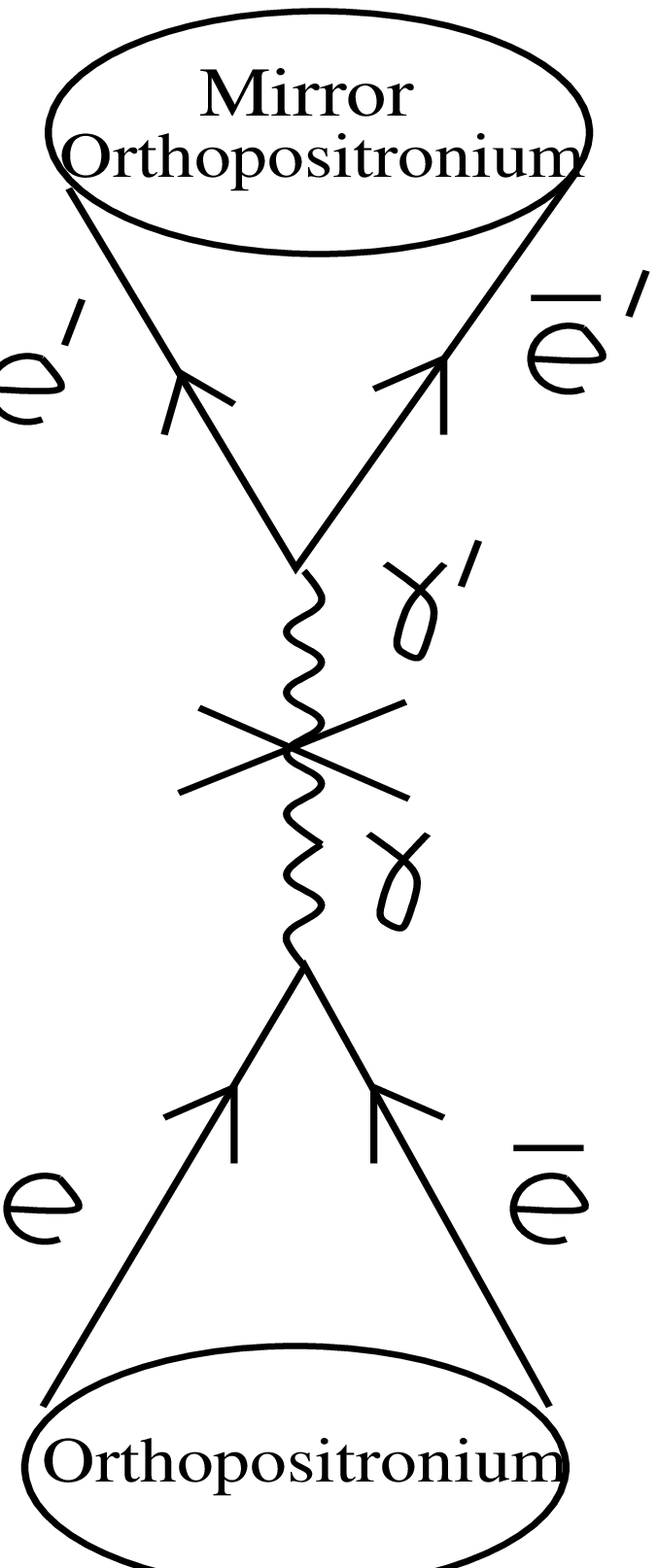,height=6.3cm,width=2.7cm}}
\vskip 0.5cm
\noindent
The effect of this mass mixing
term is to cause orthopositronium to (maximally) oscillate
into mirror orthopositronium:
\begin{eqnarray}
P(O \to O') = \sin^2 \omega t,
\end{eqnarray}
where $\omega = \pi \epsilon f$, where $f = 8.7 \times 10^4$ MHz
is the contribution to the ortho-para splitting from the one photon
annihilation diagram involving orthopositronium.

In an experiment, mirror orthopositronium decays are not 
detected, which means that
the number of orthopositronium, $N$, satisfies\cite{gl}
\begin{eqnarray}
N = \cos^2 \omega t e^{-\Gamma^{SM}t} \approx exp[-t(\Gamma^{SM} +
\omega^2 t)]
\end{eqnarray}
where $\Gamma^{SM}\simeq 7.03998\ \mu s^{-1}$ is the standard 
model orthopositronium decay rate\cite{review}.
Evidently, the observational effect of the oscillations is to
increase the apparent decay rate of ordinary orthopositronium:
$\Gamma^{eff} \approx \Gamma^{SM} + \omega^2/\Gamma^{SM}$.
In practice, orthopositronium is not produced in vacuum,
but undergoes elastic collisions at a rate, $\Gamma_{coll}$, which
depends on the particular experiment. 
These collisions cause decoherence, disrupting the oscillations.
In the limit $\Gamma_{coll} \to \infty$, the mirror world
effect goes to zero\cite{sergei}.
In all of the
existing experiments, the collision rate exceeds the
orthopositronium decay rate, which means that the apparent
decay rate is given by\cite{gn}:
\begin{eqnarray}
\Gamma^{eff} \approx \Gamma^{SM} \left(1 + {2\omega^2 \over
\Gamma^{SM}\Gamma_{coll}}\right).
\end{eqnarray}
The 1990 vacuum cavity experiment performed by a team
at the University of Michigan\cite{vac} showed a small but
statistically significant excess (about 0.1\%), which
suggested\cite{gn} an $|\epsilon |\approx 10^{-6}$. However,
a new vacuum cavity experiment\cite{vac2}, also performed
by the Michigan group, finds no anomaly:
\begin{eqnarray}
\Gamma^{exp}/\Gamma^{SM} = 1.00006 \pm  0.00018 .
\end{eqnarray}
In the 2003 experiment, the orthopositronium typically makes two wall
collisions per lifetime,
which is comparable to the 1990 Michigan experiment.
The net effect is a $2\sigma$ upper limit on the value of $\epsilon$ of:
\begin{eqnarray}
{2\omega^2 \over \Gamma^{SM}\Gamma_{coll}} < 0.00042 
\ \ \Rightarrow \ |\epsilon | &<& 5\times 10^{-7} .
\end{eqnarray}
Orthopositronium experiments can also directly search for
invisible decay modes. This can be done by tagging the positrons and
searching for events with missing energy\cite{serg}.
This would essentially be a mirror orthopositronium
`appearance' experiment (rather than a `disappearance'
experiment, which is what you get from orthopositronium
lifetime studies).
With that technique the sensitivity to photon-mirror
photon kinetic mixing can be greatly enhanced - if
the experiment is done in vacuum. This would be
very important because such an experiment
(already planned\cite{web})
could potentially probe $\epsilon$ values down to $10^{-8}$ and
possibly even lower. This would be very useful because 
there are interesting indications for
$\epsilon$ of order $10^{-8}$ 
coming from the DAMA/NaI dark matter
experiment\cite{dama2},
as we will now discuss.

If mirror matter is identified with the dark matter
in the Universe, then it is natural to interpret the dark
matter halo of our galaxy as containing mirror stars/planets/dust
and gas. 
\footnote{
According to the MACHO gravitational microlensing study\cite{macho}, 
the proportion of halo dark matter in our galaxy in compact form is in
the range $8 - 50\%$ ($95\%$ C.L.). These compact halo objects can be
interpreted as mirror stars, mirror white dwarfs etc\cite{ms}, 
with the remaining portion of the halo  ($\stackrel{>}{\sim} 50\%$) 
in the form of gas and dust.}. 
In fact, viewed from afar, by a mirror observer, our
galaxy may well resemble an elliptical galaxy --
the ordinary matter in the disk would be invisible of course.
The important point is that if the dark matter halo
of our galaxy is composed of mirror matter, then galactic mirror
atoms and dust particles can potentially
be detected in dark matter experiments via the nuclear recoil
signature\cite{f03}

The reason is that the photon-mirror photon
kinetic mixing interaction, Eq.(\ref{km}), 
gives the mirror nucleus, with (mirror) atomic number $Z'$, 
a small effective ordinary
electric charge of $\epsilon Z'e$. This means that
ordinary and mirror nuclei can elastically scatter
off each other (essentially Rutherford scattering).
For a mirror atom of mass $M_{A'}$ and (mirror) atomic number 
$Z'$ scattering
on an ordinary target atom of mass $M_{A}$ and atomic number $Z$, 
the cross section is given by:
\begin{eqnarray}
{d \sigma \over dE_R} = {\lambda \over E_R^2 v^2},
\end{eqnarray}
where $\lambda 
\equiv 2\pi \epsilon^2 \alpha^2 Z^2 Z'^2/M_A$.
In this equation,
$v$ is the velocity of the mirror nucleus in the lab frame
(i.e. where the ordinary nucleus is at rest) $E_R$ is the
recoil energy of the ordinary nucleus. 
The basic Feynman diagram for this process is given in
the following figure:
\vskip 0.4cm
\centerline{\epsfig{file=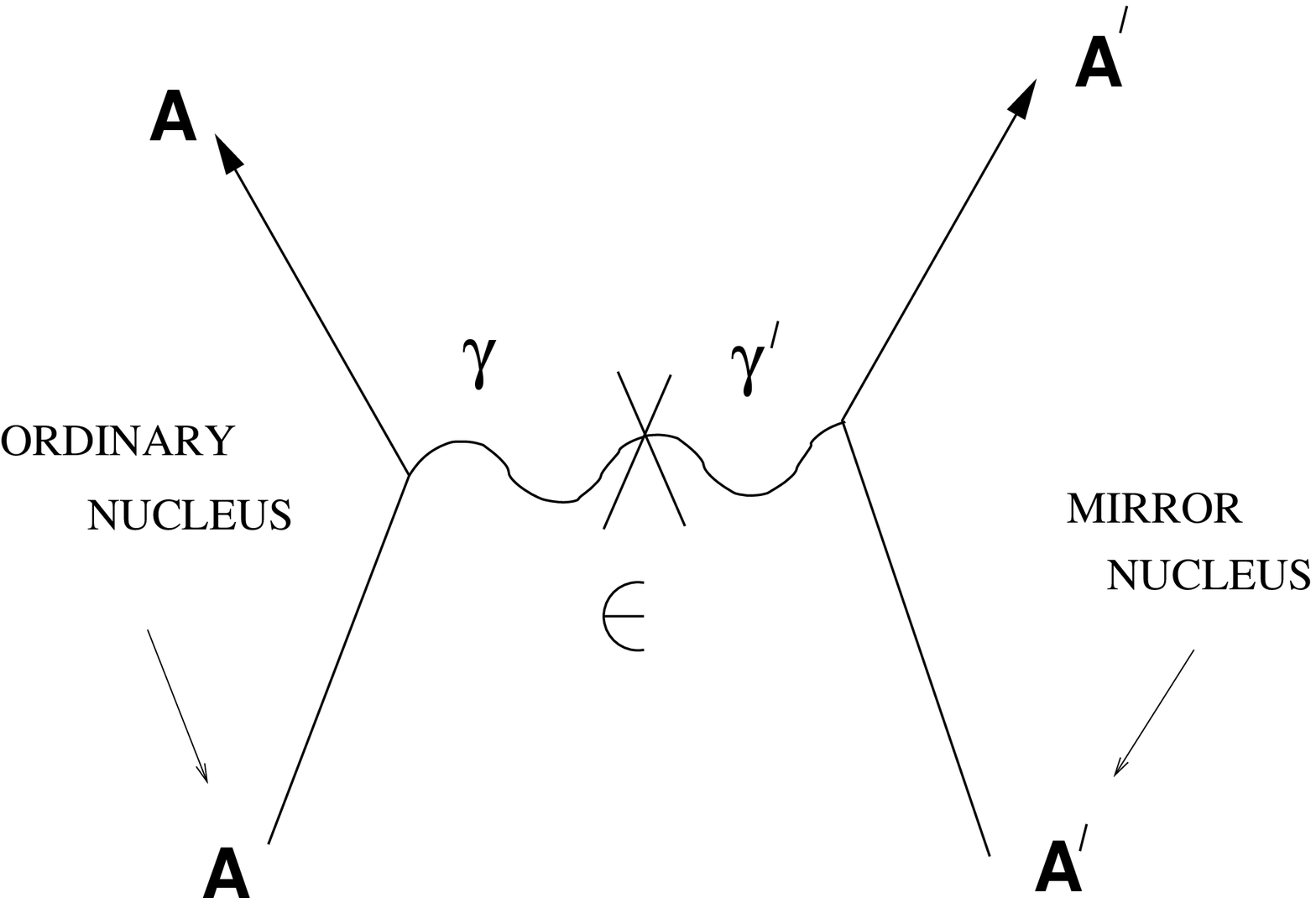,height=4.5cm,width=7.3cm}}
\vskip 0.4cm

The experiment with the most data is the DAMA/NaI
experiment\cite{dama2}.
The aim of the DAMA/NaI experiment is to measure the
nuclear recoils of Na, I atoms due to the interactions
of dark matter particles. This interaction rate should 
experience a small annual modulation due
to the Earth's motion around the sun: 
\begin{eqnarray}
A\cos 2\pi (t - t_0)/T .
\label{1}
\end{eqnarray}
According to the DAMA analysis\cite{dama2}, they indeed find 
such a modulation
over 7 annual cycles at more than $6\sigma$ C.L. Their data fit 
gives $T = (1.00 \pm 0.01)$ year and 
$t_0 = 144 \pm 22$, consistent with the expected
values. [The expected value for $t_0$ is 152 (2 June), where the
Earth's velocity, $v_E$, reaches
a maximum with respect to the galaxy].  
The strength of their signal is $A = (0.020 \pm 0.003)$
cpd/kg/keV [cpd $\equiv$ counts per day].

These are extremely impressive results which demand
serious consideration. No systematic uncertainty which
can mimic this effect has been identified and it
therefore seems probable that DAMA has discovered
dark matter.
Interestingly the interpretation of the DAMA/NaI
signal in terms of standard WIMPs appears to be
disfavoured by a number of experiments\cite{cdms,zeplin}. However, if
we interpret the DAMA/NaI signal in terms of mirror matter-type
dark matter
then the conflict with the other experiments is alleviated\cite{f03}.

In the case of a halo composed of $H', He',$ heavier
mirror elements and dust particles,
there are important differences to the standard WIMP
case due to mirror particle self interactions.
For example, assuming a number density of $n_{He'} \sim 0.08 \ cm^{-3}$
(which is suggested if $He'$ makes a significant contribution
to the halo dark matter)
the mean distance between $He' - He'$ collisions is
$1/(n_{He'} \sigma_{elastic}) \sim 0.03$ light years
(using $\sigma_{elastic} \sim 3\times 10^{-16}\ cm^2$).
One effect of the self interactions is to locally thermally equilibrate
the mirror particles in the halo.
The $He'$ (and other mirror particles) should be well described by
a Maxwellian velocity distribution with no cutoff velocity.
[$He'$ do not escape from the halo because of their
self interactions].
A temperature, $T$, common to all the mirror particles
in the halo can be defined, where
$T = M_{A'}v_0^2/2$ (of course, $T$ will depend on the spatial
position).
One effect of this is that $v_0$ 
should depend on $M_{A'}$ with $v_0 (A') = v_0 (He')
\sqrt{M_{He'}/M_{A'}}$.
Thus, knowledge of $v_0$ for $He'$ will fix $v_0$
for the other elements.
It is natural to set $v_0 (He') \sim 230$ km/s (the
sun's velocity relative to the galactic center) if the matter
density of the galactic halo is dominated by $He'$ (as 
hinted by BBN\cite{ber}).

In an experiment such as DAMA/NaI, the measured quantity is the
recoil energy, $E_R$, of a target atom. The minimum velocity
of a mirror atom of mass $M_{A'}$ impacting on a target atom
of mass $M_A$ is related to $E_R$ via the kinematic relation:
\begin{eqnarray}
v_{min} = \sqrt{ {(M_A + M_{A'})^2 E_R \over 2M_A M^2_{A'}}
} .
\label{v}
\end{eqnarray}
Values for $v_{min}$ for impacting mirror $H', He', O', Fe'$
(which span the range of interest),
for various experiments are given in the above table.
As the table shows, the experiments most sensitive to 
mirror elements are DAMA/NaI and CRESST/Sapphire, because the
other experiments have $v_{min} \gg v_0 (A')$.
Furthermore, DAMA is mainly sensitive to $O', Fe'$ and
fairly insensitive to $H', He'$. CRESST on the
other hand is sensitive to $He', O'$ and $Fe'$ (but
the CRESST experiment has much less data than the DAMA
experiment).

\begin{table}
\begin{center}
\begin{tabular}{|l|l|l|l|l|}\hline
Experiment/Target & 
$v_{min}$ ($H'$) & 
$v_{min}$ ($He'$) & 
$v_{min}$ ($O'$) & 
$v_{min}$ ($Fe'$)  
\\\hline

CRESST\cite{cresst}/$Al_2 O_3$ & 
(Al) \ 913 \ km/s & 252\ km/s & 88\ km/s & 48\ km/s
\\
 & (O) \ 720 \ km/s & 212 \ km/s & 85 \ km/s & 54 \ km/s
\\\hline
DAMA/NaI \cite{dama2} & (Na) 2834 \ km/s & 795 \ km/s & 290\ km/s & 167 \ km/s
\\
 & (I) 11656 \ km/s & 2982 \ km/s & 813\ km/s & 298 \ km/s
\\\hline
CDMS/Ge \cite{cdms}&  5992 \ km/s & 1559 \ km/s & 450\ km/s & 186 \ km/s
\\\hline
Zeplin I/Xe \cite{zeplin} &  7980 \ km/s & 2040 \ km/s & 555\ km/s & 201 \ km/s
\\ \hline

\end{tabular}
\end{center}
\end{table}

The interaction rate of halo dark matter with
a detector depends on the Earth's velocity relative
to the halo.
Because of the Earth's annual motion, its velocity
satisfies:
\begin{eqnarray}
v_E (t) &=& v_{\odot} + v_{\oplus} \cos\gamma \cos \omega (t-t_0)
\nonumber \\
&=& v_{\odot} + \Delta v_E \cos \omega (t-t_0)
\end{eqnarray}
where $v_{\odot} \approx 230$ km/s is the Sun's velocity with respect
to the galactic halo and $v_{\oplus} \simeq 30$ km/s is the Earth's
orbital velocity around the Sun (and $\omega = 2\pi/T$, with $T = 1$
year). The inclination of the Earth's orbital
plane relative to the galactic plane is $\gamma = 60^o$, which
means that $\Delta v_E \approx 15$ km/s.
The event rate in an experiment will thus contain 
an annual modulation term:
\begin{eqnarray}
R_i = R^0_i + R^1_i \cos\omega (t-t_0)
\end{eqnarray}
where
\begin{eqnarray}
R^0_i &= & {1 \over \Delta E} \int_{E_i}^{E_i+\Delta E}
\left( {dR \over dE_R}\right)_{v_E = v_{\odot}}
dE_R
\nonumber \\
R^1_i &\simeq & {1 \over \Delta E} \int_{E_i}^{E_i+\Delta E}
{\partial \over \partial v_E}
\left( {dR \over dE_R}\right)_{v_E = v_{\odot}}
\Delta v_E
dE_R\ .
\end{eqnarray}
The DAMA/NaI collaboration have found such
an annual modulation for the 2-6 keV energy
range: $R^1 (2-6 keV) = 0.020 \pm 0.003$ cpd/kg/keV.

The DAMA/NaI experiment is not very sensitive to light mirror
elements $H'$ and $He'$. The reason is the relatively high value
for $v_{min}$ (see the earlier table).
However, the DAMA/NaI experiment is quite sensitive
to any $O'$ and/or $Fe'$ component.
Interpreting the DAMA experiment in terms of mirror
$O'$, $Fe'$ mixture, the annual modulation effect
in the 2-6 keV window
can be explained if\cite{f03}:
\begin{eqnarray}
|\epsilon | \sqrt{{\xi_{O'} \over 0.10} +
{\xi_{Fe'} \over 0.02}} 
\simeq 4.5 \times 10^{-9}
\label{dama55}
\end{eqnarray}
where $\xi_{A'} \equiv \rho_{A'}/(0.3 \ {\rm GeV/cm^3})$ is the $A'$
proportion (by mass) of the halo dark matter.
The relative contribution of $O'$ and $Fe'$ can
in principle be determined by the detailed differential
spectrum in keV bins rather than using the 4 keV
window (2-6 keV).
This is illustrated in the figure below.
In this figure the solid line is
$\xi_{O'} = 0.10$, $\xi_{Fe'} \ll
0.02$, the dashed line is $\xi_{Fe'} = 0.02$, $\xi_{O'} \ll 0.10$,
and the dotted line is a 50-50 mixture, $\xi_{O'} = 0.05$,
$\xi_{Fe'} = 0.01$.
[$\epsilon = 4.5\times 10^{-9}$ in each case].  
\vskip 0.0cm
\centerline{\epsfig{file=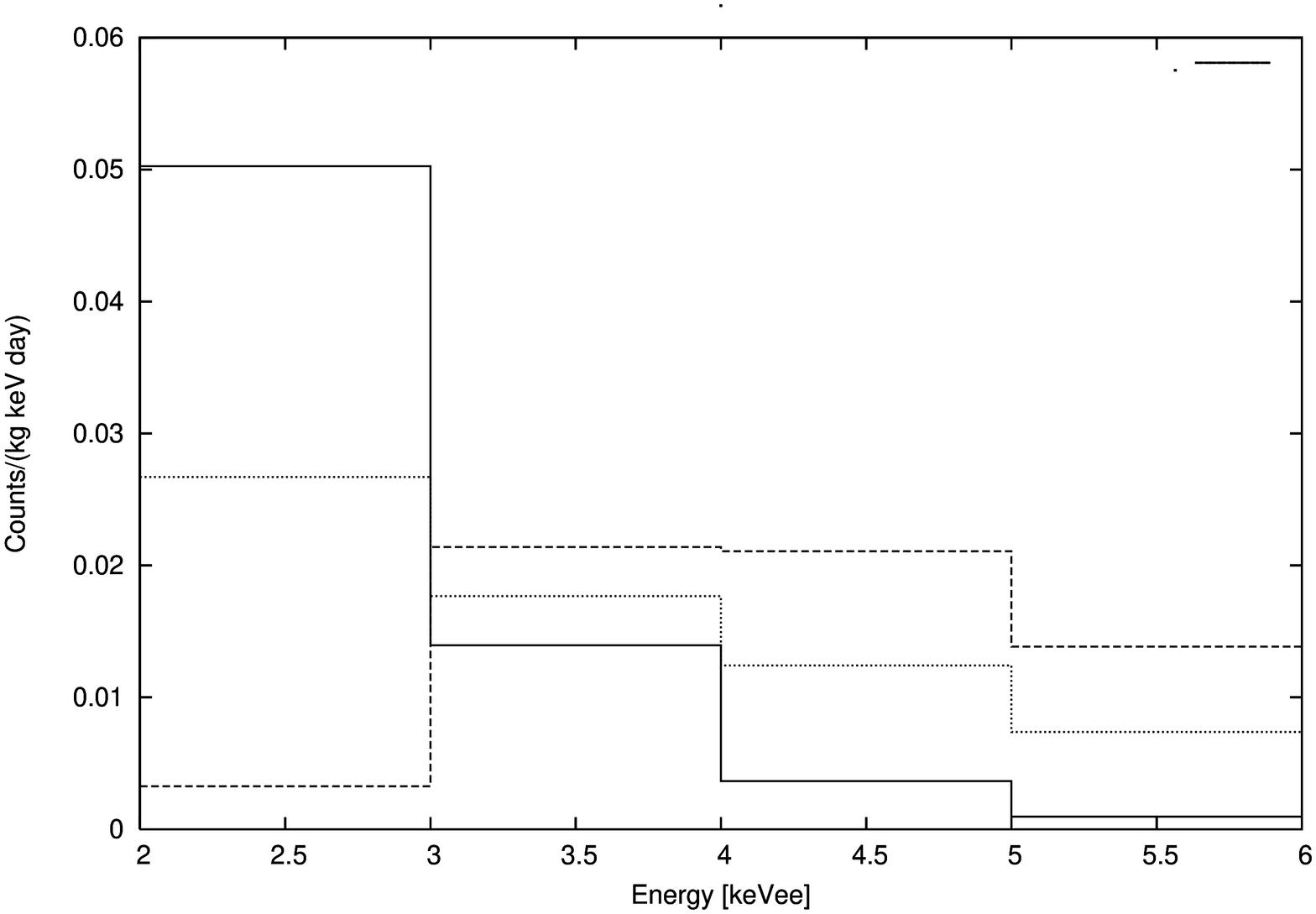, angle = 270, width=9.1cm}}
\vskip 0.3cm

\noindent
This interpretation appears to be consistent with other 
experiments, in contrast to the standard WIMP 
interpretation of the DAMA/NaI signal.

Of course there are many other implications of
mirror matter-type dark matter. Perhaps
the most fascinating possibility is that
our solar system contains mirror matter
space-bodies\cite{sil,tung}. Collisions of such bodies with themselves and
ordinary bodies would generate a population of
dust particles and larger bodies which could
impact with the Earth. The impact velocity must
be in the range:
\begin{eqnarray}
11\ km/s \stackrel{<}{\sim} v \stackrel{<}{\sim} 70 \ km/s
\end{eqnarray}
Small dust particles could be detectable in
simple surface experiments. In particular, experiments such
as the St. Petersburg experiment\cite{drob} are sensitive to
solar system mirror dust particles\cite{sm03}. Such particles can
produce a burst of bremsstrahlung photons upon passing
through ordinary matter. These photons can be detected via
a PM tube,
and the velocity of the mirror dust particle thereby determined.
Ordinary cosmic rays should be travelling close to the
speed of light, and can thereby be distinguished from
relatively slow moving mirror dust particles.
The St. Petersburg experiment finds a positive signal corresponding
to a flux of about 1 mirror dust particle per square meter per day.

Impacts of larger bodies should be less frequent, nevertheless there
is a fascinating range of evidence for their existence.
The largest recorded impact event was the 1908 Tunguska
event. Remarkably no asteroid or cometary remnants 
were recovered from the Tunguska site\cite{rev}. People have
{\it assumed} that the impacting body was made of ordinary
matter, however there is no solid evidence to support
this claim.
The Tunguska body may have been made
out of dark matter -- which is a logical possibility
if mirror matter is identified with the dark 
matter of the Universe. In fact, this hypothesis seems
to provide a better explanation for the known features
of the Tunguska event\cite{tung}.
There are also many other `anomalous' impact events, on smaller
scales\cite{small}, and evidence for anomalous impact events on
larger scales\cite{haines} which seem to be explicable if interpreted
as mirror matter impacts.
Other solar system evidence for mirror matter
also exists coming from the lack of
small craters on the asteroid EROS\cite{eros,s1} and also
from the anomalous slow-down of {\it both}
Pioneer spacecraft\cite{anderson,v2}. This overall situation is
summarized in the figure below:
\vskip 0.2cm
\centerline{\epsfig{file=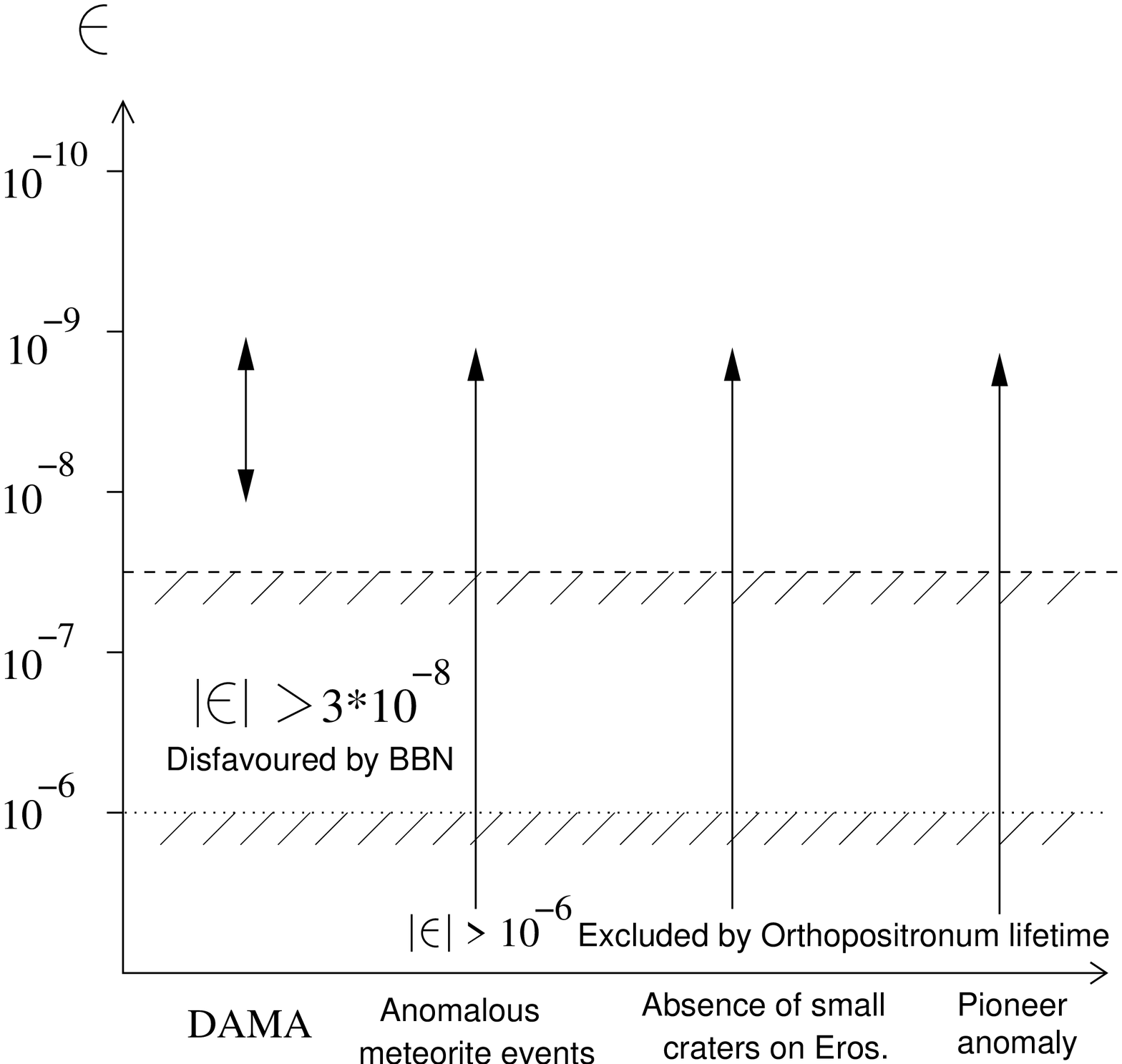,height=6.0cm,width=8.0cm}}
\vskip 0.2cm

Finally, let us mention that another large impact event has occurred 
recently in Siberia, devastating about 100 square kilometers of
forest\cite{tf} (c.f. $\sim 2100$ square kilometers in
the 1908 Tunguska event\cite{rev}).  Preliminary searches have not
found any meteorite fragments, despite the existence
of a large number of small craters at the site\cite{tf}.
If this impact event is due to a (pure) mirror matter body, it should
not have slowed down as rapidly in the atmosphere 
as an ordinary matter body
(for $\epsilon \sim 10^{-8}-10^{-9}$ as suggested by DAMA/NaI
results, the air molecules typically pass through the
body losing only a relatively small fraction of their
momentum\cite{s1}).
This might be testable from satellite observations
of the bolide (which are obviously unavailable for the
1908 Tunguska event, but should be available for this
recent Siberian event)\footnote{Of course, it is also possible
that the body could have a small amount of ordinary matter
embedded within it which would make it `opaque' to air 
molecules. However, even in this case there will be important
differences due to the rate of ablation of the body: For an ordinary
matter body, the surface heats up rapidly and is continuously melting 
thereby reducing the size of the body. In the mirror matter case, the
heat is spread out within the entire volume of the body (not just
on the surface), which
means that the rate of ablation is much lower than in the ordinary
matter case. Since smaller bodies decelerate more quickly
than larger bodies, the reduced ablation rate for
impacting mirror matter objects will imply a reduced rate of
deceleration in the atmosphere.}.
Direct detection of mirror matter fragments in the ground is also
possible at
these impact sites. The 
photon-mirror photon kinetic mixing interaction can lead to a small
static force which can keep small mirror matter fragments
near the Earth's surface\cite{cent}.
Such fragments can be experimentally detected via
the centrifuge technique\cite{cent}
and through the thermal effects of the embedded mirror matter on
the surrounding ordinary matter\cite{th}.

In conclusion, mirror matter-type dark matter is
a well motivated alternative to standard WIMP
dark matter. In fact, mirror matter-type dark matter
seems to be theoretically preferred since it requires
only a single hypothesis - mirror symmetry of fundamental
interactions. In comparison, the preferred WIMP
models require at least three {\it independent} hypothesis a) low
energy supersymmetry b) 
the lightest susy particle (LSP) is neutral and
c) R-parity exists (to keep the LSP thing stable).
Of course, the 
important point is that experiments can in principle
test the mirror matter dark matter hypothesis, and there is
currently ($> 6$ sigma!) evidence from the DAMA/NaI experiment,
along with a set of other, independent, observations which seem to
support the mirror matter-type dark matter hypothesis.


\section*{Acknowledgements}
I have great pleasure to thank my colleagues, 
S. Gninenko, A. Yu. Ignatiev, H. Lew, S. Mitra, Z. Silagadze, R. Volkas,
and T. Yoon,
for their valuable collaboration, and also to
S. Gninenko, for inviting me to contribute to this
important workshop.




\end{document}